\documentclass[preprint,pra,aps,superscriptaddress,prd]{revtex4}
\usepackage{epsfig,amsmath,graphicx,amssymb}
\usepackage[colorlinks=fals,linkcolor=blue]{hyperref}
\def\be{\begin{equation}}
\def\ee{\end{equation}}
\def\bea{\begin{eqnarray}}
\def\eea{\end{eqnarray}}

\begin{document}

\title{Periodic-cylinder vesicle with minimal energy}
\author{Xiaohua Zhou}
\email{xhzhou08@gmail.com} \affiliation{Department of Mathematics
and Physics, Fourth Military Medical University, Xi'an 710032,
China}
\date{\today}

\begin{abstract}
We give some details about the periodic cylindrical solution found
by Zhang and Ou-Yang in [Phys. Rev. E 53, 4206(1996)] for the
general shape equation of vesicle. Three different kinds of periodic
cylindrical surfaces and a special closed cylindrical surface are
obtained. Using the elliptic functions contained in
\emph{mathematic}, we find that this periodic shape has the minimal
total energy for one period when the period-amplitude ratio
$\beta\simeq1.477$, and point out that it is a discontinuous
deformation between plane and this periodic shape. Our results also
are suitable for DNA and multi-walled carbon nanotubes (MWNTs).


\textbf{PACC:} 8720, 8725D, 8745
\end{abstract}

 \maketitle


\section{Introduction}
Lipid bilayer membranes have abundant shapes in aqueous environment.
The equilibrium shape of a vesicle is determined by the minimization
of the Helfrich free energy:$^{[1]}$
$F_H=\int\int[\frac{k_{c}}{2}(2H+C_{0})^{2}+\bar{k}\Lambda] dA$.
Where $H,~\Lambda$ and $C_0$ are the mean curvature, Gaussian
curvature and spontaneous curvature, respectively, $dA$ is the
surface area element, and $k_{c}$ and $\bar{k}$ are constants. For
close shapes, we need the surface constraint and volume constraint:

$F_S=\lambda\oint dA$ and $F_V=\Delta P\oint dV$, respectively. Here
$dV$ is the volume element for the vesicle, $\lambda$ and $\Delta P$
are constants. The total energy is $F=F_H+F_S+F_V$. By studying the
first variation of the total energy $\delta^{(1)}F=0$, Ou-yang and
Helfrich obtained the general shape equation of closed
membranes$^{[2,3]}$
\bea\label{general equation}
\nonumber
              k_c(2H+C_0)(2H^2-2\Lambda-C_0H)& &\\
    -2\lambda H+2k_c\nabla^2H+\Delta P &=& 0,
\eea
where the $\nabla^{2}$ is the Laplace-Beltrami operator.
Eq.(\ref{general equation}) is a high order nonlinear partial
differential equation. It is a hard task to solve it analytically.
Only the general solution for cylindrical vesicles has been
discussed$^{ [4,5]}$ and other special solutions including the
Clifford torus,$^{[6]}$ the discounts,$^{ [7]}$ and the
beyond-Delaunay surface$^{[8]}$ have been given. Since the numeric
methods, such as finite element method using the \emph{Surface
Evolver},$^{[9]}$ were used to search solutions for this equation,
many interesting shapes were obtained in Refs.[10-18]. Compared with
those complicate shapes obtained by numeric methods, present known
analytical solutions are only the tip of iceberg in the abundant
solutions of Eq.(\ref{general equation}). In addition, a generalized
model with an arbitrary energy density function
$\mathcal{F}=\mathcal{F}(H,\Lambda)$ was considered and the
corresponding equilibrium equation was obtained.$^{[19]}$ Moreover,
in Refs.[20-22], Tu and Ou-Yang obtained the basic equations for
open vesicle with free edges using the exterior differential method.
These basic equations are also complicated and we can suppose that
they also have abundant solutions.

\section{Periodic-cylinder vesicle with minimal energy}

\begin{figure}
\includegraphics[scale =0.4]{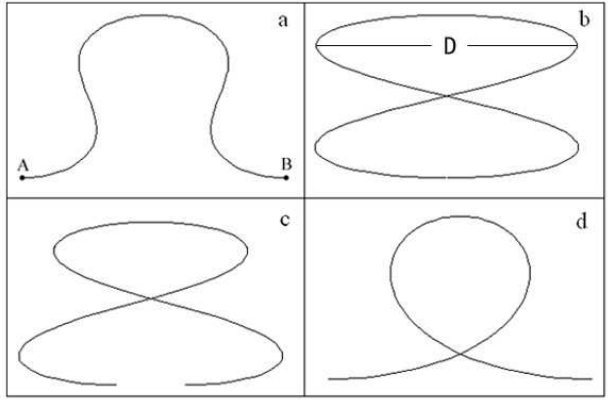}
\caption{\label{fig1}Four kinds of different shapes obtained by
choosing different $\alpha$ in expression (\ref{s}). (a) A period of
a kind of periodic cylinder when $0<\alpha<0.462$; (b) A special
closed cylinder when $\alpha=0.652$, the width-high ratio
$D/(2X_m)\simeq0.328$; (c) A period for a new kind of
self-intersected periodic cylinder when $0.652<\alpha<1$; (d) A
period for a kind of self-intersected periodic cylinder when
$\alpha>1$.}
\end{figure}

\begin{figure}
\includegraphics[scale =0.8]{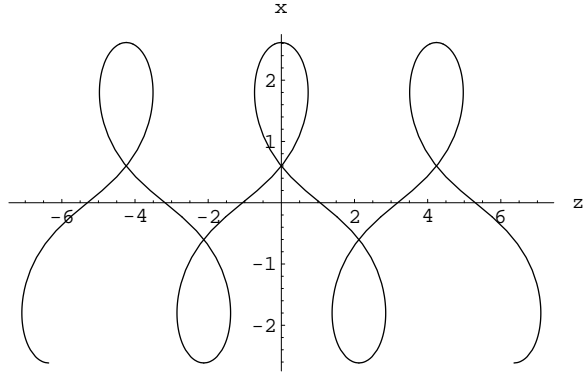}
\caption{\label{fig2}A kind of periodic shape (three periods) by
choosing $\alpha=0.9$ in expression (\ref{s}). This shape and the
shape in Fig.2 of Ref.[4] are close to the experimental shapes in
Fig.1 of Ref.[25].}
\end{figure}

In cylindrical case, Eq.(\ref{general equation}) is reduced to an
elliptical equation which was fully discussed in Ref.[5]. Let a
cylindrical surface along the $y$ axis and
$\tan\psi(x)=\frac{dz}{dx}$, here $\psi(x)$ is the angle between the
$x$ axis and the tangent to the curve at point $x$, the mean
curvature and Gaussian curvature of this surface are
\bea
H=\frac{1}{2}\cos\psi\frac{d\psi}{dx}=\frac{1}{2}\frac{d(\sin\psi)}{dx},~\Lambda=0.
\eea
Besides the planar solution $d \psi/dx=0$, the shape equation is
reduced to
\bea\label{shape}
\frac{d^2g}{d\psi^2}-4\tan\psi\frac{d
g}{d\psi}-(1-2\tan^2\psi)g=\frac{\sec^2\psi}{x_0^2},
\eea
where $g=(\frac{d\psi}{dx})^2$ and $x_0$ defined by$
\frac{1}{2x_0^2}=\frac{\lambda}{k_c}+\frac{C_0^2}{2}$. Here we let
$x_0=\sqrt{\frac{k_c}{k_c C_0^2+2\lambda}}$. Furthermore, elastic
theory is also used to analysis the shape of DNA chain and MWNTs. In
planar case, total energy of DNA$^{[23]}$ and MWNT$^{[24]}$ can be
written as $F=\frac{k_c}{2}\int(K-C_0)^2ds+\lambda\int ds$ ($K$ is
the curvature of the central line of DNA and MWNT and $ds$ is the
element of arc length), which induces the same shape equation as
Eq.(\ref{shape}). Zhang and Ou-Yang found a solution for
Eq.(\ref{shape}) with the form$^{[4]}$
\bea\label{s1}
  \sin\psi=\frac{1}{4\alpha}\left(\frac{x}{x_0}+C_1\right)^2-\alpha,
\eea
here $\alpha$ and $C_1$ are constants and we choose $C_1=0$ because
it is corresponding to a remove along $x$ axis. Choosing the reduced
value $X=x/x_0$, expression (\ref{s1}) can be written as
\bea\label{s}
  \sin\psi=\frac{1}{4\alpha}X^2-\alpha.
\eea
The cross section of this cylinder can be obtained by
\bea\label{z}
Z(X)-Z(X_1)=\int_{X_1}^X \tan\psi dX,
\eea
where $Z=z/x_0$. Choosing different $\alpha$, we find that this
solution can give us four kinds of different shapes as been shown in
Fig.\ref{fig1}. Here we only discuss the case $\alpha>0$, because we
will obtain the similar results when $\alpha<0$. Valid shapes are
only in the range $0<\alpha<0.462$. When $\alpha\geq0.462$, the
shape will be self-intersected. In Fig.\ref{fig2}, we give a new
kind of periodic shape using Fig.\ref{fig1}(c). Specifically, when
$\alpha\simeq0.652$, the points $A$ and $B$ in Fig.\ref{fig1}(a)
will be superposed and we get a closed cylinder with the width-high
ratio $D/(2X_m)\simeq0.328$ in Fig.\ref{fig1}(b). Simultaneously we
need
\bea
               & & -X_m\leq X\leq X_m~for~0<\alpha<1,\\
               & & -X_m\leq X\leq X_n~or~X_n\leq X\leq
               X_m~for~\alpha>1,~~
\eea
where $X_m=2\sqrt{\alpha(\alpha+1)}$ and
$X_n=2\sqrt{\alpha(\alpha-1)}$.

\begin{figure}
\includegraphics[scale =0.8]{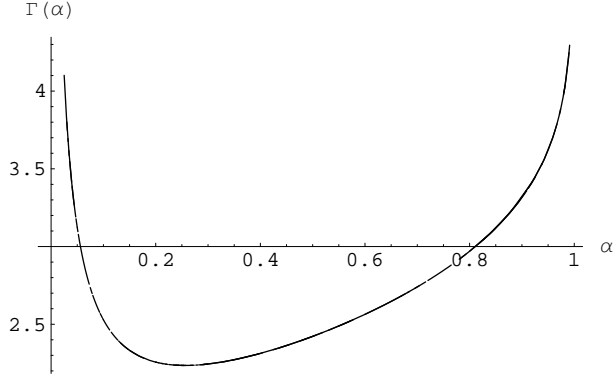}
\caption{\label{fig3}The curve of $\Gamma(\alpha)$. It has the
minimum $\Gamma_{min}\simeq2.235$ at $\alpha\simeq0.256$. When
$\alpha\rightarrow0$ and $\alpha\rightarrow1$, we have
$\Gamma\rightarrow\infty$.}
\end{figure}

Now, we turn to discuss the total energy of this solution. For
simplicity, we only consider the case $0<\alpha<1$. Then, the total
energy (actually the energy density along $y$ axis) for a period is
\bea\label{f1}
\nonumber
    \mathcal{F}_1(\alpha) &=& x_0\int_{X_m}^{-X_m}\frac{k_c(2H+C_0)^2+2\lambda}{\cos\psi}dX\\
                          &=& 4\sqrt{k_c^2 C_0^2+2\lambda k_c}\Gamma(\alpha),
\eea
here we define
\bea
    \theta={i~\rm arcsinh}\sqrt{\frac{1+\alpha}{1-\alpha}},~\textrm{k}=\frac{\alpha-1}{\alpha+1},
\eea
the function $\Gamma(\alpha)$ is
\bea
      \Gamma(\alpha) &=&
      \frac{i}{\sqrt{\alpha(\alpha+1)}}\Big\{(\alpha+1)\times\rm{E}[\theta,k]-(2\alpha+1)\times\rm{F}[\theta,k]\Big\},
\eea
where \rm{F}[x,y] and \rm{E}[x,y] are the elliptic integral of the
first kind and the seconde kind, respectively. We find that
$\Gamma(\alpha)$ has the minimum $\Gamma_{min}\simeq2.235$ at
$\alpha\simeq0.256$ (see Fig.\ref{fig3}). Thus, the minimal total
energy for one period is
\bea\label{f_min}
    \mathcal{F}_{min} =8.94\sqrt{k_c^2 C_0^2+2\lambda k_c}.
\eea
Sense it is difficult to discuss the stability of this infinite
periodic solution, we only can think that the stable shapes satisfy
$\alpha\simeq0.256$.

\begin{figure}
\includegraphics[scale =0.8]{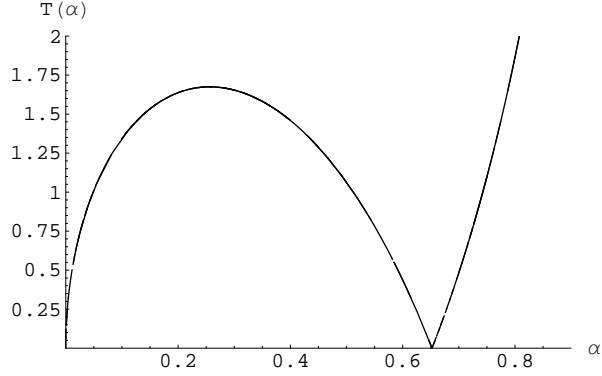}
\caption{\label{fig4}The curve of reduced period $T(\alpha)$. It has
the maximum (for not self-intersected shapes) $T_{max}\simeq1.672$
at $\alpha\simeq0.256$. When $\alpha\rightarrow0$, we have
$T\rightarrow0$.}
\end{figure}

The reduced period $T$ is
\bea\label{T}
              T(\alpha) &=& 2\int_{-X_m}^{X_m} \tan\psi dX =  \frac{8i \alpha}{\sqrt{\alpha(\alpha+1)}}\bigg\{(\alpha+1)\times \rm{E}[\theta,k]-\rm{F}[\theta,k]\bigg\}.
\eea
We show $T(\alpha)$ in Fig.\ref{fig4} and find that $T(\alpha)$ has
the maximum $T_{max}\simeq1.672$ at $\alpha\simeq0.256$. Here we
should note that the real period which can be measured in experiment
is $T'=x_0 T$ and we have $T'_{max}\simeq1.672x_0$. Further, it is
interesting to know why $\mathcal{F}_1(\alpha)$ and $T(\alpha)$ have
the minimum and maximum when $\alpha\simeq0.256$, respectively.
However, because the elliptic function is everywhere continuous but
everywhere non-derivable, it is difficult to discuss this
coincidence.

\begin{figure}
\includegraphics[scale =0.8]{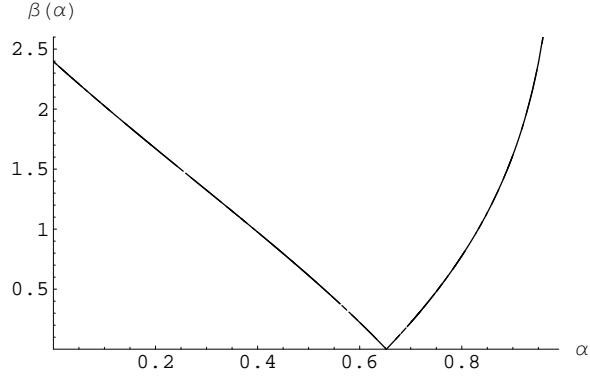}
\caption{\label{fig5}The curve of period-amplitude ratio
$\beta(\alpha)$. It has the maximum (for not self-intersected
shapes) $\beta_{max}\simeq2.4$ when $\alpha\rightarrow0$, and
$\beta(0.256)\simeq1.477$.}
\end{figure}

An important characteristic parameter for this periodic shape is the
period-amplitude ratio, it is
\bea\label{beta}
    \beta (\alpha) &=& \frac{T(\alpha)}{X_m}=\frac{4i}{\alpha+1}\Big\{(\alpha+1)\times \rm{E}[\theta,k]-\rm{F}[\theta,k]\Big\}.
\eea
We find that $\beta(\alpha)$ has the maximum (for not
self-intersected shapes) $\beta_{max}\simeq2.4$ when
$\alpha\rightarrow 0$ (see Fig.\ref{fig5}), which is in accord with
the value in Ref.[4]. Specially, when $\alpha=0.256$, we get
$\beta=1.477$. Through measuring the shapes in Ref.[26], we obtain
$\beta\approx2$ for Fig.9, $\beta\approx3.3$ for Fig.11(a), and
$\beta\approx2.7$ for Fig.11(b) and (c). It indicates that not all
these shapes can be explained by this solution. Moreover, in
Ref.[26] the authors found that planar surface can gradually change
into periodic cylinder. But in this solution, it seems that the
planar surface cannot change into periodic cylinder continuously,
because planar surface means $\beta\rightarrow\infty$ which is out
of the range $\beta<2.4$. Considering planar surface is always
stable scampered with any periodic surface,$^{[27]}$ we think there
is an energy barrier between this two kinds of surfaces. Supposing
that a plane with $\tilde{C}_0$ and $\tilde{\lambda}$ changes into a
periodic cylinder with $C_0$ and $\lambda$, then the energy change
between this deformation can be obtained by the following way.

\begin{figure}
\includegraphics[scale =0.8]{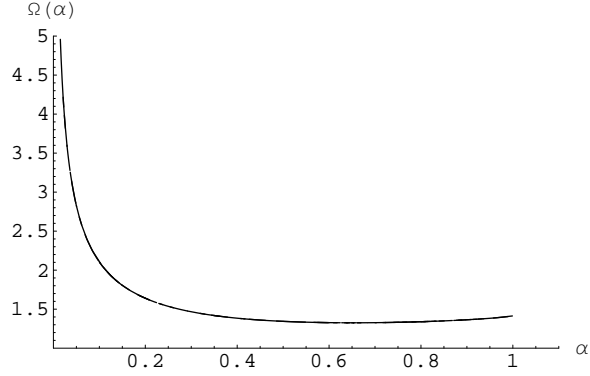}
\caption{\label{fig6}The curve of $\Omega(\alpha)$. It has the
minimum $\Omega_{min}\simeq1.325$ when $\alpha\simeq0.652$. When
$\alpha\rightarrow0$ we have $\Omega\rightarrow\infty$ and when
$\alpha\rightarrow1$ we get $\Omega\simeq\sqrt{2}$. In addition, we
have $\Omega(0.256)\simeq1.526$, $\Omega(0.462)\simeq1.356$.}
\end{figure}

For this periodic cylinder, the length for one period of the cross
section line is
\bea\label{length}
    L_1(\alpha) = 2x_0 \int_{-X_m}^{X_m} \frac{1}{\cos\psi} dX=-i~8 x_0\sqrt{\frac{\alpha}{\alpha+1}}\times \rm{F}[\theta,k].
\eea
Consider the energy density for planar surface solution is
$\mathcal{F}_{p}=\frac{k_c}{2}\tilde{C}_0^2+\tilde{\lambda}$, the
energy change for a period is
\bea\label{d1}
 \nonumber    \Delta\mathcal{F}_{p} &=& \mathcal{F}_{1}(\alpha)-\mathcal{F}_{p}L_1(\alpha)\\
                                    &=& \mathcal{F}_{1}(\alpha)-(\frac{k_c}{2}\tilde{C}_0^2+\tilde{\lambda})L_1(\alpha).
\eea
Let $\tilde{C}_0=C_0$ and $\tilde{\lambda}=\lambda$, there is
\bea\label{d11}
 \nonumber    \Delta\mathcal{F}_{p} = 4\sqrt{k_c^2 C_0^2+2\lambda k_c}\Omega(\alpha),
\eea
with
\bea\label{II}
    \Omega(\alpha)=\Gamma(\alpha)+i~\sqrt{\frac{\alpha}{\alpha+1}}\times\rm{F}[\theta,k].
\eea
 We show
$\Omega(\alpha)$ in Fig.\ref{fig6} and find that $\Omega(\alpha)$
has a minimum $\Omega_{min}\simeq1.325$ at $\alpha\simeq0.652$,
which is corresponding to the closed shape in Fig.1(b). Simply let
the barrier be equal to the energy difference, then the minima
barrier is between the planar and the self-intersected 8-shape in
Fig.\ref{fig1}(b). Clearly, in phase space, if we want that a planer
drop to a periodic cylinder, we need $ \Delta\mathcal{F}_{p}<0$ in
Eq.(\ref{d1}). It cannot be satisfied if $\tilde{C}_0=C_0$ and
$\tilde{\lambda}=\lambda$, because $\Omega(\alpha)>0$. Thus, in this
downward transition process, $\lambda$ (or $C_0$) will change.

When $\alpha>1$ we find that $\mathcal{F}_1(\alpha)$ is monotone
decreasing following the increase of $\alpha$ and
$\mathcal{F}_1=4\pi k_c C_0$ when
 $\alpha\rightarrow\infty$. Another way, when $\Delta P=0$, a circular
  cylinder solution with radii $R$ needs $C_0R=1$.$^{[28]}$ Its total energy is
$\mathcal{F}=\oint \frac{k_c}{2}(1/R+C_0)ds=4\pi k_c C_0$ which is
in accordance with the above result (note that
$\alpha\rightarrow\infty$ is corresponding to a circular
cylinder$^{[4]}$).

Remarkably, although the shape in Fig.\ref{fig1}(b) is
self-intersected for 2D surface, it is useful to describe the DNA
and MWNT shapes, because that any perturbations on the binormal
direction will probably lead it to be not self-intersected. A ring
solution: $x=R\sin\psi$ for Eq.(\ref{shape}) induces
\bea\label{ring1}
\tilde{\lambda} R^2+\frac{1}{2}k_c(\tilde{C}_0^2R^2-1)=0.
\eea
If a DNA ring changes into the shape in Fig.\ref{fig1}(b) and
$\tilde{C}_0$ and $\tilde{\lambda}$ change to $C_0$ and $\lambda$,
respectively, we have
\bea\label{ring2}
2\pi R=L_1(0.652).
\eea
Eqs.(\ref{ring1}) and (\ref{ring2}) yield the relationship
\bea\label{ring-relation}
    2.846\times(\tilde{C}_0^2+2 \tilde{\lambda}/k_c)=C_0^2+2 \lambda/k_c.
\eea
If $\tilde{C}_0=C_0=0$, expression (\ref{ring-relation}) indicates
that the tension $\lambda$ will increase in this deformation.

\section{Discussion and conclusion}

In conclusion, we have studied the particulars about the periodic
cylinder in Ref.[4] for closed shape equation of vesicle. Several
characteristics for this periodic shape are obtained, which give us
some interesting details about this solution. For instance, when the
period-amplitude ratio $\beta\simeq1.477$, this periodic shape has
minimal total energy for one period. We hope the shape with
period-amplitude ratio $\beta\simeq1.477$ can be found in
experiment. As for the other periodic surface: beyond-Delaunay
surface,$^{[8]}$ because there isn't analytic integral function of
its total energy, we don't know whether there is similar result.

Finally, we point out an interesting revelation. We know that if the
energy density of an one-dimensional structure, such as DNA$^{[23]}$
and MWNTs,$^{[24]}$ can be written as
\bea\label{ED-DNA}
\mathcal{F}=\frac{1}2{}k_c(K-C_0)^2,
\eea
 the corresponding shape equation in planar case is
\bea\label{Eq-DNA}
2\ddot{K}+K^3-(C_0^2+2\lambda/k_c)K=0,
\eea
where $\dot{K}=dK/ds$. This equation is equal to Eq.(\ref{shape})
(except the straight line solution $K=0$). The shapes of DNA and
MWNT in planar case satisfy this equation. Thus, the solutions of
Eq.(\ref{shape}) have broad meaning. Ou-Yang \emph{et al.} proved
that the energy density of MWNTs can be written as
$\mathcal{F}=\alpha K^2$,$^{[24]}$ but this model seems not suitable
for single-walled carbon nanotubes (SWNTs), because the constant
$\alpha$ will be zero for SWNTs. However, we note that the shape in
Fig.2 of Ref.[4] and the new shape in Fig.\ref{fig2} of this paper
seems to be in good agreement with the experimental SWNT shapes in
Fig.1 of Ref.[25]. Thus, we suppose that the energy density of SWNTs
satisfies expression (\ref{ED-DNA}) under suitable approximation. We
will try to prove this supposition in our future work.

This work is partly supported by National Science Foundation for
Young Scientists of China, Grant No. 10804129.

\end{document}